\documentclass[twocolumn,superscriptaddress,floatfix,preprintnumbers,amssymb ,amsmath]{revtex4}
\usepackage{graphicx}
\usepackage{dcolumn}
\usepackage{bm}
\usepackage[latin1]{inputenc}
\usepackage[mathscr]{eucal}
\usepackage{epsfig}
\begin{document}
\title{Dissipated work and fluctuation relations for non-equilibrium single-electron transitions}
\author{Jukka P.\ Pekola}
\affiliation{Low Temperature Laboratory (OVLL), Aalto University School of Science, P.O. Box 13500, 00076 Aalto, Finland}
\author{Aki Kutvonen}
\affiliation{COMP CoE at the Department of Applied Physics, Aalto University School of Science,
P.O. Box 11000, 00076 Aalto, Finland}
\author{Tapio Ala-Nissila}
\affiliation{COMP CoE at the Department of Applied Physics, Aalto University School of Science,
P.O. Box 11000, 00076 Aalto, Finland}
\affiliation{Department of Physics, Brown University, Providence RI 02912-1843}

\begin{abstract}
We discuss a simple but experimentally realistic model system, a single-electron box (SEB),
where common fluctuation relations can be tested for driven electronic transitions.
We show analytically that when the electron system on the SEB island is overheated
by the control parameter (gate voltage) drive,
the common fluctuation relation (Jarzynski equality) is only approximately valid due to dissipated heat
even when the system starts at thermal equilibrium and returns to it after the drive has been stopped. However, an
integral fluctuation relation based on total entropy production
works also in this situation. We perform extensive Monte Carlo simulations of
single-electron transitions in the SEB setup and find good agreement with the theoretical predictions.
\end{abstract}

\date{\today}

\maketitle

Statistical mechanics of small systems has been in the focus of intense interest over the past years. The common fluctuation relations, formulated, e.g., in Refs. \cite{bochkov81,jarzynski97, crooks99,crooks00}, introduce equalities to describe irreversible processes. In the thermodynamic limit these are replaced by inequalities, of which the second law of thermodynamics is the best known one. The said equalities govern statistical averages over many repeated realizations of a given process driven by external control parameters. An individual realization is either dissipative or it extracts energy from the heat bath; however, the Jarzynski equality (JE)
\begin{equation} \label{noeq1}
\langle e^{-\beta (W-\Delta F)}\rangle =1
\end{equation}
should be valid \cite{jarzynski97}. Here $\langle \cdot\rangle$ refers to averaging over an infinite number of repetitions, 
$W$ is the work done in the driven process, $\Delta F$ is the free-energy difference between the equilibrium states of the system at the end points of the drive trajectory, and $\beta \equiv (k_BT)^{-1}$ is the inverse temperature of the heat bath. The conditions for JE to be valid are amazingly
few \cite{jarzynski97,jarzynski08}: foremost, the system needs to be in thermal equilibrium with the bath in the beginning of the drive. Typically, the validity and conditions of Eq. \eqref{noeq1} have been theoretically discussed for abstract model systems \cite{jarzynski08,campisi11}, and experimentally for systems, where controlled large sample averages are difficult to obtain \cite{ritort10}.

\begin{figure}
    \includegraphics[width=8.5cm]{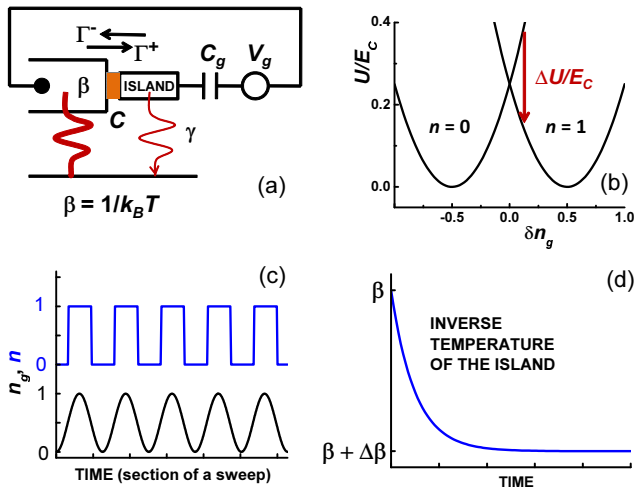}
    \caption{Single-electron box (SEB). (a) Circuit diagram showing the electronic configuration and schematically the energy relaxation by wavy lines. Electrons tunnel at rates $\Gamma^\pm$ between the isothermal lead on the left and the small island in the center. The potential of the island is controlled capacitively (capacitance $C_g$) by a variable gate voltage $V_g$. The island is not necessarily isothermal due to the dissipative gate driving and weak energy relaxation ($\gamma$) to the phonon bath. For more details, see text. (b) The energies of the two lowest lying charge states $n=0$ and $n=1$ in the gate voltage range around $\delta n_g=0$. The vertical red arrow depicts energy release in a $n:0\rightarrow 1$ transition at a positive value of $\delta n_g$. (c) A section of the harmonic drive and a schematic presentation of the corresponding transitions of charge number $n$ on the island. (d) Schematic presentation of the approximate island temperature evolution during the drive in quasi-equilibrium (regime (i), see text).}
    \label{fig1}
\end{figure}
A physical system of driven single-electron tunneling at low temperatures \cite{averin86} satisfies the requirements of concrete experimental feasibility: averaging over large number of realizations, and simple but accurate expressions of transition rates and energy relaxation become available. It has been recently shown theoretically \cite{ap11} and experimentally \cite{saira12} that Eq. \eqref{noeq1} is valid for driven isothermal transitions in a single-electron box provided detailed balance is obeyed. The general assertion is that this equality is valid even if the system is overheated by the control drive or it is driven to a full non-equilibrium state \cite{jarzynski08}. Here we demonstrate that this is not the case for the system we consider. Instead, we find that an integral fluctuation theorem due to Seifert \cite{seifert05} applies even in this situation. We expect the same conclusion to hold in general for driven, overheated systems where detailed balance is not obeyed.
%

The single-electron box (SEB) \cite{mb,sac} (cf. Fig.~\ref{fig1}) considered in Refs.~\cite{ap11,saira12,jp12}, is a simple, yet a representative system for our arguments. In a SEB, a tunnel contact admits electrons to enter or leave the island of the box. The electrostatic energy of the box with $n$ (integer number) excess electrons on the island is given by $U(n,n_g) = E_C(n-n_g)^2$. Here $E_C=e^2/2C_\Sigma$ is the elementary charging energy of the box determined by the total capacitance $C_\Sigma=C+C_g+C_0$, where $C$ is the capacitance of the tunnel junction, $C_g$ the gate capacitance and $C_0$ the self-capacitance of the island. The relative energies of the different charge states $n$ are determined by the gate voltage $V_g$ via $n_g=C_gV_g/e$. In this work we discuss dynamics in the range $0\le n_g \le 1$, and at low temperatures, $\beta E_C\gg 1$ such that the charge number can have values $n=0$ or $n=1$ only. When an electron tunnels into ($+$) or out of ($-$) the island, energy $\pm [U(0,n_g)-U(1,n_g)]=\pm 2E_C\delta n_g$ is released.
We have written  $\delta n_g \equiv n_g-1/2$. The heat bath at temperature $T$ is that of the phonons, and the lead of the SEB (on the left in Fig. \ref{fig1} (a)) is assumed to be a reservoir in equilibrium with the phonon bath. However, unlike in Refs. \cite{ap11,saira12}, here we allow the small island of the SEB to be driven into non-equilibrium as a consequence of a gate protocol with many dissipative transitions (Fig. \ref{fig1} (c)) and slow energy relaxation of the electrons.

In Ref.~\cite{ap11} it was concluded that Eq.~\eqref{noeq1} is always valid in a driven SEB if all the electrodes remain at the bath temperature $T$. This condition prevails if the drive injects non-equilibrium electrons (holes in out-tunneling) on the SEB island at a rate which is slower than the energy relaxation rate $\gamma$ of non-equilibrium excitations. The opposite limit $f \gg \gamma$, where $f$ is the frequency of the cyclic gate drive, is the domain of the present discussion. In this case there are two main regimes to consider as regards the energy distribution on the island. (i) Electron-electron relaxation rate is much faster than the drive, $\gamma_{\rm e-e} \gg f\gg \gamma$, in which case the electrons on the island occupy a Fermi-Dirac distribution with higher temperature than the bath, see Fig. \ref{fig1} (d). We emphasize that here we do not introduce two separate heat baths in the problem, but the second temperature is solely a result of the energy deposition on the island due to the drive by the control gate. If the protocol of the control parameter is stopped during the process to an arbitrary value, the system will relax to canonical equilibrium corresponding to the value of the control parameter. Thus the system obeys balance condition in the sense of, and as required by the fluctuation theorems \cite{crooks00, jarzynski11}. (ii) The injection rate is the largest frequency in the problem, $f \gg \gamma_{\rm e-e},\gamma$, which leads to a full non-equilibrium on the island, i.e., the electronic energy distribution deviates from the Fermi-Dirac one. If we consider the standard metallic SEBs, regime (i) is typical in experiments since $\gamma_{\rm e-e} \approx 10^9$ s$^{-1}$ in an ordinary metal, whereas $\gamma \approx 10^6$ s$^{-1}$ for the corresponding electron-phonon relaxation at the sub-kelvin temperatures where the SEBs are typically operated \cite{saira12,pothier97,saira10}. In this work we discuss quantitatively the case (i) only.
%
%
We consider a symmetric gate drive around the degeneracy, moving between charge states $n=0$ and $n=1$~\cite{other_trajectories} as depicted in Fig. \ref{fig1} (c). We assume unit peak-to-peak amplitude, where $\delta n_g(t)$ varies between $-1/2$ and $+1/2$. We assume as usual that the system is in equilibrium with the bath before the gate drive starts at $t=0$, as requested by the fluctuation relations in general. The drive ends after $N$ back and forth ramps of the gate. The thermodynamic work done is straightforward to obtain from its definition~\cite{jarzynski97}, or from the Markovian viewpoint~\cite{crooks00}. Since the beginning and end points of the drive are the same, $\delta n_g =-1/2$, there is no ambiguity of the proper expression of work to use~\cite{work}, and the dissipation can be written as $W-\Delta F =\sum_i \pm \Delta U_i$, where the $+$ sign refers to the rising $n_g$ half-periods, and $-$ sign to the decending ones \cite{ap11}. Moreover, for the sake of a simple argument, we assume in our analytical treatment that exactly one transition between the two charge states occurs in each half-cycle of the gate drive. This is a regime that can be achieved to a high accuracy - the probability of half-cycles with multiple transitions in a fully normal system to be considered below is $\approx 3.0\mathcal{T}^2/(\beta^4E_C^2e^4R_T^2)$, where $\mathcal{T}$ is the duration of the half-cycle and $R_T$ is the junction resistance. Numerical results will be provided later for more general cases, where also multiple jumps in each leg can occur. 
For the $i$th half-cycle of the gate drive (increasing $\delta n_g$), we may write in the single-jump approximation by standard path-averaging arguments \cite{ap11}
\begin{eqnarray} \label{noeq3}
&&\langle e^{-\beta \Delta U}\rangle= \\ &&\int_{0}^{\mathcal{T}} d\tau e^{-\beta \Delta U(\tau)} e^{-\int_{0}^\tau d\tau' \Gamma^+(\tau')}\Gamma^+(\tau) e^{-\int_{\tau}^{\mathcal{T}} d\tau' \Gamma^-(\tau')},\nonumber
\end{eqnarray}
where $\Delta U =2E_C\delta n_{g}$ is the work dissipated in the half-cycle, and $\delta n_{g}$ is the gate position where the transition occurs. $\Gamma^+$ is the transition rate into the island that depends explicitly on time due to the gate drive, but also on the island temperature that can be different from that of the bath due to the dissipative transitions in the earlier half-cycles. The tunneling rates we apply in what follows are those for static biasing conditions, since we envision drive frequencies that are far below any relevant energy scale in the system. The descending half-cycles where an electron tunnels out of the island are identical in terms of the energetics and our argument. In what follows we assign the dissipated work in a multi-leg ramp as $W-\Delta F$, and in a single leg as $\Delta U$.

If the island temperature stays constant at $T$, the detailed balance for the tunneling rates into and out of the island holds, $\Gamma^-(t)=e^{-\beta \Delta U(t)}\Gamma^+(t)$, and by simple arguments presented in the supplementary on-line material we obtain from Eq. \eqref{noeq3}
\begin{eqnarray} \label{noeq4}
\langle e^{-\beta \Delta U}\rangle=P_{{\rm R}}=1.
\end{eqnarray}
Here $P_{{\rm R}}$ is the probability of making a $n:1\rightarrow 0$ transition in the corresponding reverse path, and the last step follows from our assumption of the symmetry of the path and of exactly one transition in each leg. Since in the isothermal system, all the legs of the drive are independent, we see that JE is trivially satisfied in this case. This argument is more general in the isothermal process, not limited to the single-jump trajectories~\cite{ap11}.

The situation is qualitatively different if the island is driven out of equilibrium.
In what follows, we consider tunneling in a fully normal metal box, where
\begin{eqnarray} \label{noeq5}
\Gamma^\pm = \frac{1}{e^2R_T}\int dE f(E,\beta)[1-f_{\rm NE}(E\pm\Delta U)]
\end{eqnarray}
are the tunneling rates when the lead is in equilibrium with Fermi-Dirac distribution $f(E,\beta)$ and island has a distribution $f_{\rm NE}(E)$. For the full equilibrium case, $f_{\rm NE}(E)=f(E,\beta)$, we have $\Gamma^\pm_0 =\pm(e^2R_T)^{-1} \Delta U (1-e^{\mp\beta \Delta U})^{-1}$.  For the island in regime (i),
$f_{\rm NE}(E)=f(E,\beta+\Delta \beta)$. A further simplified analysis in this regime can be carried out if we assume that
the system stays close to equilibrium. We write $\Delta\beta$ for the instantaneous deviation of the inverse temperature of the island from that of the bath at the moment of the tunneling event,
and assume that $|\Delta \beta/\beta| \ll 1$. Linear expansion in $\Delta \beta$ then yields
\begin{eqnarray} \label{noeq6}
\Gamma^\pm \simeq \Gamma^\pm_0(1-\frac{e^2R_T}{2}\Gamma^\mp_0 \Delta \beta).
\end{eqnarray}
We see immediately that detailed balance at temperature $T$ is not obeyed any more if $\Delta \beta \neq 0$, but
\begin{eqnarray} \label{noeq6a}
\Gamma^- /\Gamma^+\simeq e^{-\beta \Delta U}[1-\frac{1}{2}\Delta U\Delta \beta].
\end{eqnarray}
We find a corresponding linear correction to the expression in Eq. \eqref{noeq3} for the $i$th half period of the gate drive as
\begin{eqnarray} \label{noeq7}
\langle e^{-\beta \Delta U}\rangle \simeq 1-\frac{1}{2} \langle Q\rangle \Delta \beta.
\end{eqnarray}
Here $\langle Q\rangle$ is the average value of heat dissipated in crossing the degeneracy, which is evaluated for the isothermal system here: according to Ref. \cite{ap11}, $\langle Q\rangle>0$ is approximately proportional to the sweep rate.

Although $\langle e^{-\beta \Delta U}\rangle \neq 1$ in Eq. \eqref{noeq7} in general, this
of course does not prove JE wrong,
since the requirement of
equilibrium at the beginning of the gate (half-)period was not imposed here. However, we may make use of Eq. \eqref{noeq7} to prove that
JE does not hold in the following simple example where the ramp starts under the equilibrium conditions.
Suppose that the symmetric gate ramp consists of only two linear legs ($N=1$), see inset in Fig. \ref{fig:2sweeps_sep}. Again exactly one tunneling event occurs in each
period. Next, we assume that the first tunneling event heats or cools the island by an amount determined by its heat capacity $\mathcal{C}$,
such that $\Delta T =0$ before the first tunneling, and $\Delta T = E/\mathcal{C}$ after the first tunneling, where
$\Delta T =-k_B^{-1}\Delta \beta/\beta^2$
and $E$ is the energy deposited on the island by the tunneling electron. Furthermore, we assume that the relaxation of heat is so slow
that the temperature of the island is changing only at the transitions during the sweep~\cite{equil}. In this sweep, we may then write
similar evolution for the full cycle as in Eq. \eqref{noeq3}, but now assuming that the temperature on the island depends on the
energy at which the electron tunnels in the first leg. This analysis is presented in the supplementary on-line material. After straightforward algebra we obtain for the dissipated work $W-\Delta F =\Delta U_1-\Delta U_2$
\begin{eqnarray} \label{noeq8}
\langle e^{-\beta(W-\Delta F)}\rangle\simeq 1 -\frac{k_B}{4\mathcal{C}}\beta^2\langle Q\rangle^2.
\end{eqnarray}

This equation shows that although $\langle e^{-\beta \Delta U_1}\rangle =1$ for the first half-period with $\Delta\beta =0$,
and $\langle e^{-\beta \Delta U_2}\rangle > 1$ for the second half-period with typically $\Delta \beta <0$, the overall average assumes values $\langle e^{-\beta(W-\Delta F)}\rangle < 1$, due to the heating induced correlation between the two legs. We note by using Jensen's inequality, $e^{- \langle x \rangle} \le \langle e^{- x} \rangle$,
that this conclusion is consistent with the second
law of thermodynamics.

To check our analytic predictions we performed numerical simulations of single electron transitions in
the SEB setup by using the
standard stochastic Monte Carlo simulation method. Unlike in the analysis above, in these simulations
no approximations were made: the number of jumps within a leg was not restricted to one, and we used the exact form of the tunneling rates instead of the linearized approximation of Eq. (\ref{noeq6}). Furthermore, in the analytical results, we have expanded the tunneling rates up to the second order in $\Delta\beta$ for more precise comparison.
We set the resolution of $\Delta U(t)\in [-1,1]$ to $0.0005$ using linear increments
and the resolution of the temperature difference between the lead and the island $\Delta T \in [0,2T]$
to $0.001T$. The control parameter protocol consisted of a linearly increasing half sweep
$n_g(t):[0,t_f],n_g(t)=t/t_f$ and a linearly decreasing one $n_g(t):[t_f,2t_f],n_g(t)=2-t/(2t_f)$, with
$4000$ steps. We set the sweep rate $f=2/t_f$ to $10^7 \text{s}^{-1}$,
which gives about 95\% single jump legs.
In each simulation the temperature of the heat bath was set to $T = 0.1$ K,
the charging energy to $E_C = 40k_B T$ and the tunneling resistance to $R_T = 450$ k$\Omega$.

\begin{figure}
    \includegraphics[width=8.5cm]{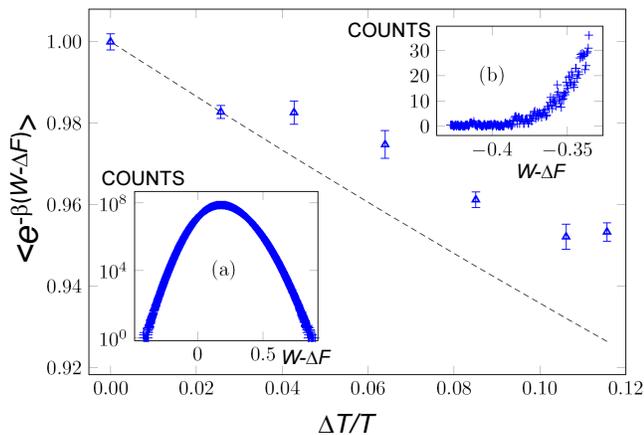}
    \caption{Simulation results (triangles) of $\langle e^{-\beta (W-\Delta F)} \rangle$ and the corresponding theoretical
approximation of the single jump and small $\Delta T/T$ approximation, Eq. (8) (dashed line). The horizontal axis shows
the average temperature after the first half sweep. The corresponding values of the heat capacities from left to
right are: $50 E_C/\text{K}$, $30 E_C/\text{K}$,
$20 E_C/\text{K}$, $15 E_C/\text{K}$, $12 E_C/\text{K}$ and $11 E_C/\text{K}$.
The error bars are the standard error of the mean of the corresponding data.
Each data point is obtained from $1.75 \times 10^{11}$ independent realizations of the tunneling process.
The insets (a) and (b) show the sampling from simulation with $\mathcal C=12 E_C/\text{K}$ with all the
$1.75 \times 10^{11}$ repetitions, (b) demonstrating explicitly that even the tails of the distribution are well sampled.}
    \label{fig:2sweeps}
\end{figure}
We calculated the tunneling rates (Eq. \eqref{noeq5})
numerically using the standard trapezoidal rule integration. We determined the integration limits
and the number of trapezoids by doubling both of them until the error was
less than $10^{-15}$ for all values of $\Delta U$ and $\Delta \beta$. The heat generation
method is presented in detail in the supplementary material. We assumed weak coupling between
the electrons on the island and the phonon bath. Thus the temperature of the island was
controlled only by the heat generation due to the tunneling events during the sweep.

To numerically test the prediction of Eq. (8) for $\langle e^{-\beta (W-\Delta F)} \rangle$, we performed
$1.75 \times 10^{11}$ repetitions for different values of the heat capacity $\mathcal{C}$ ranging from 11$E_C/\text{K}$ to 50$E_C/\text{K}$. The simulation data are shown in Fig. \ref{fig:2sweeps}. First, as expected JE is accurately
satisfied in the limit of $\Delta T=0$. This was verified with several different sweep rates (data
not shown). Most importantly, we find that
$\langle e^{-\beta (W-\Delta F)}\rangle$ decreases with increased heating
in good agreement with the theoretical prediction~\cite{multilegs}.
%

\begin{figure}
    \includegraphics[width=8.5cm]{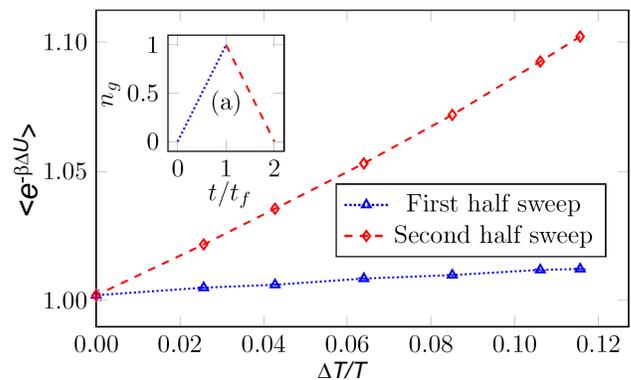}
    \caption{Simulation results of $\langle e^{-\beta \Delta U} \rangle$ over the first and the second half-sweep
as a function of the scaled average temperature after the first half sweep. The corresponding values of the
heat capacities for the data points are the same as in Fig.~\ref{fig:2sweeps}.
The starting point of the second sweep is the temperature difference produced in the corresponding first half
sweep, while the first half sweep starts from equilibrium. The statistical errors are of the size of the data points.
Each data point is obtained from $1.75 \times 10^{11}$ independent realizations of the tunneling process.
The inset (a) illustrates the control parameter protocol during the first half and the second half of the sweep.}
    \label{fig:2sweeps_sep}
\end{figure}

To explicitly demonstrate the details behind the violation of the JE,
we calculated $\langle e^{-\beta \Delta U} \rangle$ separately for the increasing and decreasing
parts of the sweep. The results are
shown in Fig. \ref{fig:2sweeps_sep}. The average is very close to unity for the first half of the
sweep, which starts in equilibrium. The slow increase is a result of the contribution from multijump legs. However, during the second half sweep the initial
temperature of the island is higher than that of the bath
and we obtain $\langle e^{-\beta \Delta U} \rangle > 1$ for that leg.
This result is in agreement with the theoretical prediction of Eq. (7).

An integrated fluctuation theorem (IFT) based on
total entropy production, including the entropy change of the system and the surrounding medium along a single trajectory, $\Delta s_{\rm tot}$, was discussed by Seifert in~\cite{seifert05}. It was concluded that IFT in the form
\begin{eqnarray} \label{mj1}
\langle e^{-\Delta s_{\rm tot}}\rangle = 1
\end{eqnarray}
applies without either the assumption of equilibrium starting condition or detailed balance. We can apply the single-trajectory entropy production, with the assumptions in the analytic treatment of this paper in a straightforward manner, and obtain the relation
\begin{eqnarray} \label{mj7}
\langle e^{-\Delta s_{\rm tot}}\rangle=\langle e^{-\beta \Delta U}\rangle - \frac{1}{2}\Delta \beta \langle \Delta Ue^{-\beta \Delta U}\rangle.
\end{eqnarray}
By applying the single-jump techniques above, we find that $\langle \Delta Ue^{-\beta \Delta U}\rangle=-\langle Q\rangle$, and inserting this result and Eq. (\ref{noeq7}) into Eq. (\ref{mj7}), we conclude that Eq.~(\ref{mj1}) is valid identically (at least within the approximations made) for a leg with arbitrary island temperature. Therefore Eq.~(\ref{mj1}) is valid also for the overheated two-leg trajectory where Eq.~\eqref{noeq1} fails.

In summary, we have analyzed the prediction of the Jarzynski equality for driven single-electron transitions under
experimentally relevant conditions: the system may be either in equilibrium, in quasi-equilibrium (overheating) or in true non-equilibrium. The JE is satisfied in the first case. We have shown both
analytically and numerically that Eq. (\ref{noeq1}) is not applicable when dissipative heat causes a temperature change
in the SEB, although the system starts at equilibrium and returns to it. However, the integrated fluctuation theorem
of Eq. (\ref{mj1}) holds even in this case. The experimental realization of our setup
requires low tunneling rates
and slow relaxation rate between the system and the bath: these requirements are best satisfied using tunnel junctions between normal metals and superconductors \cite{saira10}, or in semiconducting quantum dots \cite{utsumi10,kung12}.

Acknowledgements: This work has been supported in part by the Academy of Finland through its COMP and LTQ CoE grants. We thank Udo Seifert, Dmitri Averin, Liao Chen, Massimiliano Esposito, See-Chen Ying, Frank Hekking, Paolo Solinas, Risto Nieminen and Erik Aurell for useful discussions.


\begin{widetext}
\title{Dissipated work in non-equilibrium single-electron transitions, on-line material}

\author{Jukka P. Pekola, Aki Kutvonen, and Tapio Ala-Nissila}



\date{\today}

\maketitle
\section{Derivation of Eq. (3) of the main text}
We start with Eq. (2) of the main text:
\begin{eqnarray} \label{noeq3}
&&\langle e^{-\beta \Delta U}\rangle= \int_{0}^{\mathcal{T}} d\tau e^{-\beta \Delta U(\tau)} e^{-\int_{0}^\tau d\tau' \Gamma^+(\tau')}\Gamma^+(\tau) e^{-\int_{\tau}^{\mathcal{T}} d\tau' \Gamma^-(\tau')}.
\end{eqnarray}
Inserting the detailed balance condition $\Gamma^-(t)=e^{-\beta \Delta U(t)}\Gamma^+(t)$ into this equation we may rewrite it into
\begin{eqnarray} \label{noeq4}
&&\langle e^{-\beta \Delta U}\rangle= \int_{0}^{\mathcal{T}} d\tau  e^{-\int_{0}^\tau d\tau' \Gamma^+(\tau')}\Gamma^-(\tau) e^{-\int_{\tau}^{\mathcal{T}} d\tau' \Gamma^-(\tau')}.
\end{eqnarray}
Next we make a change of variable in the main integral, $t=\mathcal{T}-\tau$, and note that $\Gamma^\pm(u)=\Gamma^\pm_R(\mathcal{T}-u)$ for the reverse trajectory $R$ and for any time instant $u$ along the symmetric trajectories considered here, yielding
\begin{eqnarray} \label{noeq5}
&&\langle e^{-\beta \Delta U}\rangle= \int_{0}^{\mathcal{T}} dt  e^{-\int_{0}^{\mathcal{T}-t} d\tau' \Gamma^+_R(\mathcal{T}-\tau')}\Gamma^-_R(t) e^{-\int_{\mathcal{T}-t}^{\mathcal{T}} d\tau' \Gamma^-_R(\mathcal{T}-\tau')}.
\end{eqnarray}
Making the change of integration variable in the integrals of the exponents as well, $\tau''=\mathcal{T}-\tau'$, and reordering the terms in the integrand, we finally obtain
\begin{eqnarray} \label{noeq6}
&&\langle e^{-\beta \Delta U}\rangle= \int_{0}^{\mathcal{T}} dt  e^{-\int_{0}^{t} d\tau'' \Gamma^-_R(\tau'')}\Gamma^-_R(t) e^{-\int_{t}^{\mathcal{T}} d\tau'' \Gamma^+_R(\tau'')}.
\end{eqnarray}
By direct inspection we identify this as the probability of making exactly one $n:1\rightarrow 0$ transition in the reverse trajectory, i.e., $P_R$, and due to the condition that exactly one transition is taking place in its mirror trajectory, we finally conclude that
\begin{eqnarray} \label{noeq6}
&&\langle e^{-\beta \Delta U}\rangle=P_R= 1
\end{eqnarray}
in this case.

\section{Derivation of Eq. (7) of the main text}
We first expand the tunneling rate into the box
\begin{eqnarray} \label{box100}
\Gamma^+ = \frac{1}{e^2R_T}\int dE f(E,\beta)[1-f(E+\Delta U,\beta+\Delta \beta)]
\end{eqnarray}
for a small inverse temperature difference $\Delta \beta$ up to the linear correction. Since $\partial f(E,\beta)/\partial \beta = -E f(E,\beta)[1-f(E,\beta)]$, we obtain
\begin{eqnarray} \label{box101}
\Gamma^+ \simeq \frac{1}{e^2R_T}\int dE f(E,\beta)[1-f(E+\Delta U,\beta)]+ \frac{\Delta \beta}{e^2R_T} \int dE(E+\Delta U)f(E,\beta)f(E+\Delta U,\beta)[1-f(E+\Delta U,\beta)].
\end{eqnarray}
Since all the temperature arguments in Fermi distributions are now equal ($=\beta$), we may drop them for now. Equation \eqref{box101} can thus be written as
\begin{eqnarray} \label{box101a}
\Gamma^+ \simeq \frac{1}{e^2R_T}\int dE f(E)[1-f(E+\Delta U)]+ \frac{\Delta \beta}{e^2R_T} \int dE(E+\Delta U)f(E)f(E+\Delta U)[1-f(E+\Delta U)].
\end{eqnarray}
Since $f(E)[1-f(E+x)]=[f(E)-f(E+x)]/(1-e^{-\beta x})$, we obtain from Eq. \eqref{box101a}
\begin{eqnarray} \label{box102}
\Gamma^+ \simeq \Gamma^+_0 + \frac{\Delta \beta}{e^2R_T}\frac{1}{1-e^{-\beta \Delta U}} \int dE(E+\Delta U)[f(E)-f(E+\Delta U)]f(E+\Delta U),
\end{eqnarray}
where
\begin{eqnarray} \label{box102a}
\Gamma^+_0 = \frac{1}{e^2R_T}\int dE f(E,\beta)[1-f(E+\Delta U,\beta)]=\frac{1}{e^2R_T}\frac{\Delta U}{1-e^{-\beta\Delta U}}
\end{eqnarray}
is the tunneling rate into the box at equilibrium temperature.
Let us derive the analytic expression of the integral in Eq. \eqref{box102} step by step for illustration.
\begin{eqnarray} \label{box102b}
&&\int dE(E+\Delta U)[f(E)-f(E+\Delta U)]f(E+\Delta U) = \nonumber \\&&\int dE (E+\Delta U)f(E+\Delta U)[1-f(E+\Delta U)]-\int dE(E+\Delta U)f(E+\Delta U)[1-f(E)]=\nonumber \\&&\int dE'\, E' f(E')[1-f(E')]-\int dE'\,E' f(E')[1-f(E'-\Delta U)].
\end{eqnarray}
Using
\begin{eqnarray} \label{box201c}
\int dE\,E f(E)[1-f(E+x)]= -\frac{1}{2}\frac{x^2}{1-e^{-\beta x}},
\end{eqnarray}
we obtain from Eq. \eqref{box102b},
\begin{eqnarray} \label{box102d}
&&\int dE(E+\Delta U)[f(E)-f(E+\Delta U)]f(E+\Delta U) = \frac{1}{2}\frac{\Delta U^2}{1-e^{\beta \Delta U}}.
\end{eqnarray}
Combining Eqs. \eqref{box102} and \eqref{box102d}, we get
\begin{eqnarray} \label{box103}
\Gamma^+ \simeq \Gamma^+_0 - \frac{\Delta \beta}{2e^2R_T}\frac{\Delta U}{1-e^{-\beta \Delta U}} \frac{\Delta U}{e^{\beta \Delta U}-1}=\Gamma^+_0(1-\frac{e^2R_T}{2}\Gamma^-_0 \Delta \beta ),
\end{eqnarray}
where
\begin{eqnarray} \label{box103a}
\Gamma^-_0 = \frac{1}{e^2R_T}\int dE f(E)[1-f(E-\Delta U)]=\frac{1}{e^2R_T}\frac{\Delta U}{e^{\beta\Delta U}-1}
\end{eqnarray}
is the tunneling rate out from the box at equilibrium temperature.

Similarly we find for the opposite rate
\begin{eqnarray} \label{box104}
\Gamma^- \simeq \Gamma^-_0(1-\frac{e^2R_T}{2}\Gamma^+_0 \Delta \beta ).
\end{eqnarray}
Therefore, instead of the original detailed balance $\Gamma^+_0/\Gamma^-_0=e^{\beta \Delta U}$, we find
\begin{eqnarray} \label{box105}
\frac{\Gamma^+ }{\Gamma^-}\simeq \frac{\Gamma^+_0}{\Gamma^-_0}[1+\frac{e^2R_T}{2}(\Gamma^+_0-\Gamma^-_0) \Delta \beta]=e^{\beta \Delta U}[1+\frac{1}{2}\Delta U \Delta \beta],
\end{eqnarray}
which is the reason why JE fails. Here, in the last step we have used $\Gamma^+_0-\Gamma^-_0=\Delta U/(e^2R_T)$.

Now consider a symmetric trajectory of the gate from time $0$ to $\tau$. We assume that one and only one jump occurs when crossing the energy degeneracy. We find for this leg
\begin{eqnarray} \label{box106}
&&\langle e^{-\beta\Delta U}\rangle=\int_{0}^{\tau} e^{-\beta \Delta U(\tau)} e^{-\int_{0}^\tau d\tau' \Gamma^+(\tau')}\Gamma^+(\tau) e^{-\int_\tau^{T}d\tau'\Gamma^-(\tau')}d\tau \nonumber \\&& \simeq \int_{0}^{\tau}  e^{-\int_{0}^\tau d\tau' \Gamma^+(\tau')}\Gamma^-(\tau) e^{-\int_\tau^{T}d\tau'\Gamma^-(\tau')}d\tau
+\frac{1}{2}\Delta \beta\int_{0}^{\tau} \Delta U(\tau) e^{-\int_{0}^\tau d\tau' \Gamma^+(\tau')}\Gamma^-(\tau) e^{-\int_\tau^{T}d\tau'\Gamma^-(\tau')}d\tau.
\end{eqnarray}
In the second step we have made use of the result Eq. \eqref{box105}. Note that from now on we can use the expressions or values of the quantities at the initial temperature, since the corrections would yield errors in the next order only. Now, we find that the second line of Eq. \eqref{box106} can be rewritten such that
\begin{eqnarray} \label{box106b}
\langle e^{-\beta\Delta U}\rangle\simeq P_R +\frac{1}{2} \langle Q\rangle_R \Delta \beta,
\end{eqnarray}
where $P_R$ is the probability of the transition in the reverse trajectory over the same gate section, and $\langle \Delta U\rangle_R$ is the expectation value of $\Delta U$ over this reversed trajectory. Since we assume linear trajectory and the assumption of success with one jump each time $P=1$, we have $P_R =P=1$ and $\langle Q\rangle_R =-\langle Q \rangle$. Thus,
\begin{eqnarray} \label{box106c}
\langle e^{-\beta\Delta U}\rangle\simeq 1 -\frac{1}{2} \langle Q\rangle \Delta \beta.
\end{eqnarray}
We can write \eqref{box106c} also in the form
\begin{eqnarray} \label{box107}
\langle e^{-\beta\Delta U}\rangle \simeq 1-\frac{1}{2} \langle Q\rangle \Delta \beta.
\end{eqnarray}
Since $\langle Q\rangle > 0$, and $\Delta \beta <0$, we find that $\langle e^{-\beta\Delta U}\rangle >1$ in the case of an overheated island.
\section{Derivation of Eq. (8) of the main text}
Now we consider a back-and-forth gate trajectory as in the main text. We make use of Eq. \eqref{box107}
\begin{eqnarray} \label{oh1}
\langle e^{-\beta\Delta U}\rangle\simeq 1 -\frac{1}{2} \langle Q\rangle \Delta \beta
\end{eqnarray}
to prove that JE is not satisfied. Here $\beta =1/k_BT$, where $T$ is the bath temperature. We consider a simple trajectory, which consists of a linear $n_g:0\rightarrow 1$ sweep followed immediately by a similar but opposite sweep $n_g:1\rightarrow 0$. We make the simplifying assumption that in each half-period, one electron tunnels in the preferred direction. We also assume that when heat is released in the island, it adjusts its temperature instantaneously, but the relaxation rate of heat to environment is much longer than the duration of the sweep. In this sweep, lasting over a period $2\tau=\tau+\tau$, we may then write the average as
\begin{eqnarray} \label{oh2}
\langle e^{-\beta(W-\Delta F)}\rangle = &&\int_{-\infty}^\infty dE \int_\tau^{2\tau} d\tau_2 \int_0^{\tau} d\tau_1 e^{-\beta [\Delta U(\tau_1)-\Delta U(\tau_2)]}e^{-\int_0^{\tau_1} \Gamma_+(\Delta U(\tau'),T_0)d\tau'} \gamma_+(E,\tau_1)\times \nonumber\\&&e^{-\int_\tau^{\tau_2}\Gamma_-(\Delta U(\tau'),T(E))d\tau'}\Gamma_-(\Delta U(\tau_2),T(E)).
\end{eqnarray}
Unlike in the previous se section, we have dropped in the integrals the probability of the second transition, since we explicitly assume that only one transition occurs in each leg. Here, $\Gamma_\pm(\Delta U(t),T')$ are the tunneling rates into/out of the island when the island has temperature $T'$, and \begin{eqnarray} \label{oh2a}
\gamma_+(E,\tau_1)=\frac{1}{e^2R_T}f(E-\Delta U(\tau_1),\beta)[1-f(E,\beta)]
\end{eqnarray}
is the corresponding equilibrium transition rate density of $\Gamma_+$ at energy $E$. The expression in Eq. \eqref{oh2} assumes that the system starts in equilibrium (temperature $T$), and the tunneling event in the first leg influences the temperature via $T(E)=T+\Delta T(E)$, and therefore the tunneling probability out from the box in the second half of the cycle. We assume that
\begin{eqnarray} \label{oh2b}
\Delta T(E) = \frac{E}{\mathcal C},
\end{eqnarray}
i.e. the increase of temperature is proportional to the energy $E$ deposited on the island, and $\mathcal C$ is the heat capacity. Since
\begin{eqnarray} \label{oh3}
\Gamma_-(\Delta U,T')=\frac{1}{e^2R_T}\int dE f(E,\beta')[1-f(E-\Delta U,\beta)]= \frac{1}{e^2R_T}\int dE f(E+\Delta U,\beta)[1-f(E,\beta')]=\Gamma_+(-\Delta U,T')
\end{eqnarray}
always, we may write formally Eq. \eqref{oh2} as averaging over two $n_g:0 \rightarrow 1$ sweeps at the two different temperatures as
\begin{eqnarray} \label{oh4}
\langle e^{-\beta(W-\Delta F)}\rangle = &&
\int_{-\infty}^\infty dE  \int_0^{\tau} d\tau_1 e^{-\beta \Delta U(\tau_1)}e^{-\int_0^{\tau_1} \Gamma_+(\Delta U(\tau'),T)d\tau'} \gamma_+(E,\tau_1)\times \nonumber\\&&\int_0^{\tau} d\tau_2 e^{-\beta \Delta U(\tau_2)}e^{-\int_0^{\tau_2}\Gamma_+(\Delta U(\tau'),T(E))d\tau'}\Gamma_+(\Delta U(\tau_2),T(E)).
\end{eqnarray}
Here the second line is the average of Eq. \eqref{oh1} with $\Delta \beta =-\Delta T(E)/(k_B T^2)$. Thus we may write \eqref{oh4} as
\begin{eqnarray} \label{oh5}
&&\langle e^{-\beta(W-\Delta F)}\rangle \simeq \nonumber \\&&
\int_{-\infty}^\infty dE  \int_0^{\tau} d\tau_1 e^{-\beta \Delta U(\tau_1)}e^{-\int_0^{\tau_1} \Gamma_+(\Delta U(\tau'),T)d\tau'} \gamma_+(E,\tau_1)(1+\frac{k_B}{2}\frac{\beta^2 \langle Q\rangle}{\mathcal C}E)
\nonumber \\&&
=\int_0^{\tau} d\tau_1 e^{-\beta \Delta U(\tau_1)}e^{-\int_0^{\tau_1} \Gamma_+(\Delta U(\tau'),T)d\tau'} \Gamma_+(\Delta U(\tau_1),T)\nonumber\\ &&+\frac{k_B}{2}\frac{\beta^2 \langle Q\rangle}{\mathcal C} \int_0^{\tau} d\tau_1 e^{-\beta \Delta U(\tau_1)}e^{-\int_0^{\tau_1} \Gamma_+(\Delta U(\tau'),T)d\tau'}\times \nonumber \\&& \big[\int_{-\infty}^\infty dE\, E f(E-\Delta U(\tau_1),T)[1-f(E,T)]\big]  =
1+ \frac{k_B}{4}\frac{\beta^2 \langle Q\rangle}{\mathcal C}\langle \Delta U\rangle_R.
\end{eqnarray}
Here, $\langle \Delta U \rangle_R = - \langle Q\rangle$ is the average value of $\Delta U$ in the reverse trajectory, and finally
\begin{eqnarray} \label{oh3}
\langle e^{-\beta(W-\Delta F)}\rangle\simeq 1 -\frac{k_B}{4\mathcal C}\beta^2\langle Q\rangle^2.
\end{eqnarray}

\end{widetext}


\begin{thebibliography}{99}

\bibitem{bochkov81} G.\ N.\ Bochkov and Yu.\ E.\ Kuzovlev, Physica A {\bf 106}, 443 (1981).

\bibitem{jarzynski97} C.\ Jarzynski, Phys.\ Rev.\ Lett.\ {\bf 78}, 2690 (1997).

\bibitem{crooks99} G.\ E.\ Crooks, Phys.\ Rev.\ E {\bf 60}, 2721 (1999).

\bibitem{crooks00} G.\ E.\ Crooks, Phys.\ Rev.\ E {\bf 61}, 2361 (2000).


\bibitem{jarzynski08} C. Jarzynski, Eur. Phys. J. B {\bf 64}, 331 (2008).

\bibitem{campisi11} M. Campisi, P. H\"{a}nggi, and P. Talkner, Rev. Mod. Phys. {\bf 83}, 771 (2011).

\bibitem{ritort10} A. Alemany and F. Ritort, Europhysics News  {\bf 41} 2, 27 (2010).

\bibitem{seifert05} Udo Seifert, Phys. Rev. Lett. {\bf 95}, 040602 (2005).

\bibitem{averin86} D.\ V.\ Averin and K.\ K.\ Likharev, J.\ Low Temp.\ Phys.\ {\bf 62}, 345 (1986).

\bibitem{ap11} D.\ V.\ Averin and J.\ P.\ Pekola, EPL {\bf 96}, 67004 (2011).

\bibitem{saira12} O.-P. Saira, Y. Yoon, T. Tanttu, M. M\"ott\"onen, D. V. Averin, and J. P. Pekola, submitted (2012).

\bibitem{jp12} J. P. Pekola and O.-P. Saira, arXiv:1204.4623 (2012).

\bibitem{mb} M. B\"uttiker, Phys.\ Rev. B {\bf 36}, 3548 (1987).

\bibitem{sac} P. Lafarge, H. Pothier, E. R. Williams, D. Esteve, C. Urbina, and M. H. Devoret, Z.\ Phys. B {\bf 85}, 327 (1991).

\bibitem{jarzynski11} C. Jarzynski, Annual Reviews of Condensed Matter Physics {\bf 2}, 329 (2011).

\bibitem{pothier97} H. Pothier, S. Gu\'eron, Norman O. Birge, D. Esteve, and M. H. Devoret, Phys. Rev. Lett. {\bf 79}, 3490 (1997).

\bibitem{saira10} O.-P.\ Saira, M.\ M\"{o}tt\"{o}nen, V.\ F.\ Maisi, and J.\ P.\ Pekola, Phys.\ Rev.\ B {\bf 82}, 155443 (2010).

\bibitem{other_trajectories} We can show that the contributions of other possible trajectories ($n:0\rightarrow 0, 1\rightarrow 0$ or $1\rightarrow 1$) can be neglected, as they introduce expectation values that are exponentially (in $\beta E_C$) smaller than those arising from the main $n:0\rightarrow 1$ trajectories.

\bibitem{work} C.\ Jarzynski, C.\ R.\ Physique {\bf 8}, 495 (2007); J.\ M. Vilar and J.\ M. Rubi, Phys. Rev. Lett. {\bf 100}, 020601 (2008), and the following discussion.

\bibitem{equil} Yet after the gate ramp has stopped, the system returns to equilibrium due to this weak relaxation. Since no work is done after the ramp, the conditions of Jarzynski equality are satisfied.

\bibitem{multilegs} Besides the simple two-leg trajectory, numerical simulations were run for multiperiod trajectories consisting of either linear or harmonic (Fig.~\ref{fig1}) legs. The average $\langle e^{-\beta (W-\Delta F)}\rangle$ decreases monotonically with the number $N$ of periods (results not shown).

\bibitem{utsumi10} Y. Utsumi, D. S. Golubev, M. Marthaler, K. Saito, T. Fujisawa, and G. Sch\"on, Phys. Rev. B {\bf 81}, 125331 (2010).

\bibitem{kung12} B. K\"ung, C. R\"ossler, M. Beck, M. Marthaler, D. S. Golubev, Y. Utsumi, T. Ihn, and K. Ensslin, Phys. Rev. X {\bf 2}, 011001 (2012).




\end{thebibliography}
\end{document}